\documentclass[reprint,superscriptaddress, amsmath,amssymb,aps,prl]{revtex4-2}

\usepackage{graphicx}%
\usepackage{braket}
\usepackage{float}
\graphicspath{{figs/}}
\usepackage{dcolumn}%
\usepackage{bm}%
\usepackage[verbose]{placeins} 
\usepackage[colorlinks,linkcolor=red,urlcolor=blue,citecolor=blue]{hyperref}
\usepackage[normalem]{ulem}

\usepackage{xcolor}

\begin{document}

\preprint{APS/123-QED}

\title{Quantum feedback control with a transformer neural network architecture}
\author{Pranav Vaidhyanathan}
\affiliation{
Department of Engineering Science, University of Oxford, Oxford OX1 3PJ, United Kingdom
}

\author{Florian Marquardt}%
\affiliation{
 Max Planck Institute for the Science of Light, Staudtstr. 2, 91058 Erlangen, Germany
}
\affiliation{Department of Physics, Friedrich-Alexander-Universit\"at Erlangen-N\"urnberg, 91058 Erlangen, Germany}
\author{Mark T. Mitchison}%
 \email{mark.mitchison@kcl.ac.uk}
\affiliation{School of Physics, Trinity College Dublin, College Green, Dublin 2, D02 K8N4, Ireland}
\affiliation{Department of Physics, King’s College London, Strand, London, WC2R 2LS, United Kingdom}

\author{Natalia Ares}%
 \email{natalia.ares@eng.ox.ac.uk}
\affiliation{
Department of Engineering Science, University of Oxford, Oxford OX1 3PJ, United Kingdom
}

\begin{abstract}

Attention-based neural networks such as transformers have revolutionized various fields such as natural language processing, genomics, and vision. Here, we demonstrate the use of transformers for quantum feedback control through both a supervised and reinforcement learning approach. In particular, due to the transformer's ability to capture long-range temporal correlations and training efficiency, we show that it can surpass some of the limitations of previous control approaches, e.g.~those based on recurrent neural networks trained using a similar approach or policy based reinforcement learning. We numerically show, for the example of state stabilization of a two-level system, that our bespoke transformer architecture can achieve near unit fidelity to a target state in a short time even in the presence of inefficient measurement and Hamiltonian perturbations that were not included in the training set as well as the control of non-Markovian systems. We also demonstrate that our transformer can perform energy minimization of non-integrable many-body quantum systems when trained for reinforcement learning tasks. Our approach can be used for quantum error correction, fast control of quantum states in the presence of colored noise, as well as real-time tuning, and characterization of quantum devices.

\end{abstract}

\maketitle

\begin{figure*}
\includegraphics[width=\textwidth]{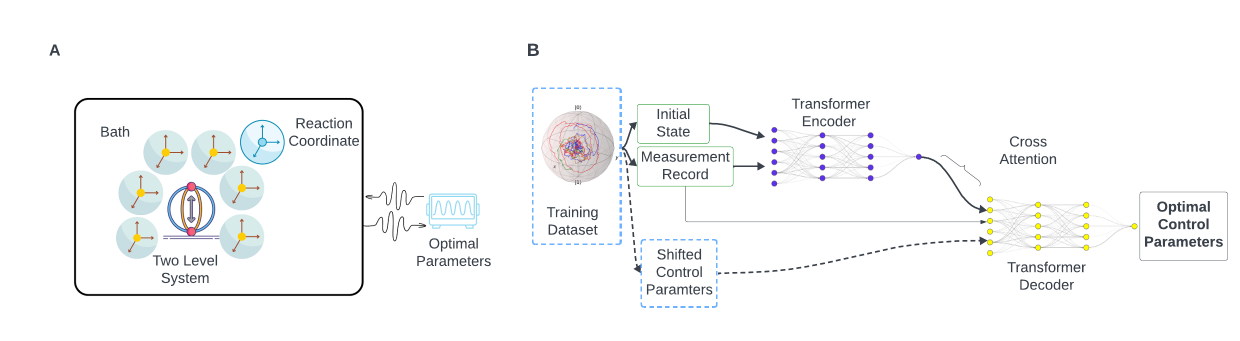}
    \caption{\label{fig:wide}\textbf{Problem and Architecture Overview:} \textbf{a,} The two level system (TLS) interacting with a bath in a non-Markovian manner that is embedded with a reaction coordinate (RC) which interacts in a Markovian manner. The measurement record obtained continously is used to predict optimal values of the control parameters by the transformer. \textbf{b,} The transformer's structure consists of an encoder and decoder architecture. During training, the encoder takes the initial state and the measurement record as input (green dotted boxes). The decoder takes the encoder output as part of the cross attention layer and the optimal parameters (blue dotted boxes), to autoregressively predict the optimal parameters for the next time steps. However, during inference, only the initial state and measurement record is given as input to the encoder (solid arrows). The decoder then predicts the optimal next values of the control parameters based on this data.}
    \label{dataset-and-setup}
\end{figure*}

\textit{Introduction.---}Quantum technologies depend crucially on our ability to precisely control quantum systems. Measurement-based feedback is an especially powerful approach to quantum control, which lies at the heart of quantum error correction and has myriad applications in the preparation and stabilization of quantum states in the presence of noise~\cite{zhang2017quantum}\cite{puviani2023boostinggottesmankitaevpreskillquantumerror}\cite{Wiseman_2002}. However, unlike noisy feedback control in the classical regime, quantum feedback faces an additional obstacle: only partial information on the quantum state is available, even in principle, due to the inherently disturbing nature of quantum measurements. In general, therefore, optimizing control fields requires --- in addition to the measurement record --- an estimate of the quantum state to be explicitly computed from some model of the dynamics~\cite{Wiseman2009}, adding overhead (e.g., extra memory or time costs) to the feedback loop.

Machine learning algorithms offer a promising route to solve this problem. Appropriately trained neural networks can provide a compact representation of the most important correlations between data, allowing for significantly more efficient feedback protocols, in principle. Recent work in this direction has demonstrated the power of both model-free~\cite{Fosel2018,Wang2020, Sivak2022,Reuer2023} and model-based~\cite{Borah2021,Porotti2023} reinforcement learning for quantum feedback control. Although the former approach is a ``black box'' that is flexible enough to be applied to a range of scenarios, the latter can exploit the physics of the system to improve efficiency. 

However, approaches using recurrent neural networks do not scale well with long-range dependencies, such as an extensive measurement record. They also suffer from the problem of vanishing gradients~\cite{bengio1994learning}. This is due to the assumption that the hidden state within each recurrent unit encodes dependencies from the previous state. This adds an inherent Markovian inductive bias that is unsuitable for processes with memory \cite{PRXQuantum.2.040355}. Clearly, this poses a challenge for feedback control of non-Markovian open system dynamics~\cite{Breuer2016}, which arises naturally in many platforms. Yet, even for Markovian open quantum systems (i.e., those described by a Lindblad equation), the measurement record is a non-Markovian stochastic process~\cite{Landi2024} because measurement backaction causes the future evolution of the state to depend unavoidably on past measurement outcomes.

In recent years, transformers and attention-based models that were originally used to model natural language \cite{vaswani2017attention}, have emerged as extremely versatile tools for various fields, ranging from genomics to robotics \cite{kamathtrans}. Due to the attention mechanism that encodes correlation between all aspects of the given input sequence, they have far outclassed recurrent neural networks (RNNs) and long-short term memory type RNNs (LSTMs) in various tasks. Recently, transformer-based approaches have also demonstrated their ability not only to adapt to model-based and model-free reinforcement learning tasks \cite{janner2021offline}\cite{chen2021decision} but also to perform on par with state-of-the-art approaches for these tasks. As highlighted by Chen \textit{et al.} these causally masked transformers simply output optimal actions and eliminate the requirement to fit value functions or calculate policy gradients\cite{chen2021decision}.

In this work, we aim to utilize the ``unreasonable" effectiveness of the attention mechanism in order to perform closed-loop feedback control for continuously measured open quantum systems that also undergo evolution due to measurement back-action~\cite{jacobs2006straightforward,Wiseman2009}. We demonstrate that an attention-based approach to quantum feedback control, with neural networks trained using both supervised and reinforcement learning, offers a robust and scalable solution that outperforms traditional methods used for quantum feedback control. Using transfer learning, we can demonstrate that the transformer also generalizes well to non-Markovian open quantum systems, which has yet to be demonstrated by existing methods.

\textit{Setup.---}We consider a quantum system undergoing a continuous weak measurement of the diffusive kind, e.g.~homodyne readout in quantum optics~\cite{Wiseman1993,Wiseman1994} or electronic charge detection by a quantum point contact~\cite{Korotkov1999,Goan2001}. Let $\hat{\rho}_t$ denote the state of the system at time $t$, conditioned on the measurement record $\mathbf{r}_t$. An experimenter uses their knowledge of the measurement record to control the system by manipulating some control parameter $\lambda_t$ entering its Hamiltonian $\hat{H}(\lambda_t)$. The conditional dynamics is then described by a stochastic master equation of the form~\cite{Wiseman2009}
\begin{equation}
	\label{SME}
d \hat{\rho}_t = \frac{1}{i\hbar} [\hat{H}(\lambda_t), \hat{\rho}_t]dt + \mathcal{D}[\hat{c}] \hat{\rho}_t dt + \sqrt{\eta} \mathcal{H}[\hat{c}] \hat{\rho}_t d W_t,
\end{equation}
where the jump operator $c$ describes the effect of coupling to the measuring device, the dissipation superoperator is $\mathcal{D}[\hat{c}] \hat{\rho}=\hat{c} \hat{\rho} \hat{c}^{\dagger}-\frac{1}{2}\left(\hat{c}^{\dagger} \hat{c} \hat{\rho}+\hat{\rho} \hat{c}^{\dagger} \hat{c}\right)$, the innovation superoperator is $\mathcal{H}[\hat{c}] \hat{\rho}= \hat{c} \hat{\rho}+\hat{\rho} \hat{c}^{\dagger}-\operatorname{Tr}\left[\left(\hat{c}+\hat{c}^{\dagger}\right) \hat{\rho}\right] \hat{\rho} $, the measurement efficiency is $\eta$, and the measurement noise in each small time step $dt$ is described by independent Wiener increments $dW_t$ with zero mean and variance $dW_t^2 = dt$~\cite{Jiang2020}. The measurement record $r_t$ increments according to
\begin{equation}
	\label{measurement_record}
	d r_t =   \operatorname{Tr}\left [ \left (\hat{c}+\hat{c}^\dagger\right ) \hat{\rho}_t \right ]  d t + \frac{d W_t}{\sqrt{\eta}}.
\end{equation}
Note that we use boldface notation to distinguish the history of the measurement record up to time $t$, $\mathbf{r}_t = (\cdots, r_{t-2dt}, r_{t-dt}, r_{t})$, from its instantaneous value, $r_t$. For simplicity, we consider a single jump operator $c$ and control parameter $\lambda_t$, but our method can be generalized straightforwardly to the case of multiple jump operators and control parameters.

In a general feedback protocol, the control parameter for the next time step is determined by the entire past history of the measurement record, i.e., it is a functional $\lambda_{t+dt}[\mathbf{r}_t]$. In the simplest case of linear feedback~\cite{Wiseman1993}, the control parameter is proportional to the measurement result, $\lambda_{t+dt}\propto dr_t$, but this approach permits a very limited class of protocols and also suffers badly from measurement inefficiencies~\cite{Wang2001,Mitchison2021a}. More general state-based methods~\cite{Zhang2020} decide the optimal feedback using a (implicit or explicit) model of the conditional quantum state, e.g.~by solving Eq.~\eqref{SME} using experimentally obtained values for the measurement noise $dW_t$. Alternatively, reinforcement learning creates an implicit model of the dynamics in terms of a probability distribution (policy function) $\pi_\theta(\lambda_{t+dt}|\mathbf{r}_t)$, which is represented as a neural network parametrized by some weights and biases $\theta$. While recent successes of this approach have been demonstrated using the RNN architecture~\cite{Fosel2018,Porotti2023}, here we take a different approach based on the transformer architecture~\cite{vaswani2017attention}.

\textit{Transformer model.---}Our model consists of a custom transformer encoder-decoder architecture (see Fig.~\ref{dataset-and-setup}), which we name QuantumEncoder and QuantumDecoder, to determine $\lambda_t$ at each time step. At its core, a transformer is designed to process sequential data by capturing long-range dependencies and contextual information. Unlike traditional recurrent neural networks (RNNs) that process sequences step by step, transformers employ a mechanism called self-attention to attend to different parts of the input simultaneously. 

The QuantumEncoder processes the initial quantum state and the measurement record, embedding it into a higher-dimensional space and capturing dependencies through self-attention mechanisms. The QuantumDecoder takes the measurement record, with positional embeddings, as its input since this detailed sequential or spatial information is not provided by the encoder's compressed representation of core features from its input (its latent space).
By feeding the measurement record into the decoder, we allow the model to adaptively adjust the optimal control parameter at each time step based on the observed system dynamics. The decoder employs self-attention that is causally masked, preventing it from `seeing' future measurements, to ensure that predictions for the optimal control parameters at each time step are based only on current and past measurements. 
During training, the QuantumDecoder module takes the optimal parameter values and measurement record as input and learns to predict the next value $\lambda_t$ in the sequence.
We use the fidelity between the evolved state due to the $\lambda_t$ and target state as the loss function used to train our model. The output of the decoder's last layer is passed through a linear transformation followed by a softmax function to obtain a probability distribution over the $\lambda_t$ values. We also set up a sweep to optimize for optimal number of layers, the learning rate, optimizer and the number of epochs (see Supplemental Material \cite{supplement}).

\textit{State stabilization in the two-level system.---}To demonstrate the effectiveness of the transformer-based approach, we showcase a numerical example of quantum state stabilization. The loss function in this case is the infidelity between the conditional state and some pure target state $\ket{\psi_{\rm targ}}$,
\begin{equation}
	\label{cost_function}
	L = 1-\braket{\psi_{\rm targ}|\hat{\rho}_t|\psi_{\rm targ}}.
\end{equation}
Our model system comprises a two-level system (TLS) with Hamiltonian 
\begin{equation}  
	\label{TLS_Hamiltonian}
	\hat{H}(\lambda_t) =\frac{\hbar \varepsilon}{2} \hat{\sigma}_z+\frac{\hbar \lambda_t}{2} \hat{\sigma}_x,
\end{equation}
undergoing a continuous measurement described by the jump operator $\hat{c} = \sqrt{\kappa}\hat{\sigma}_-$. Here, $\varepsilon$ denotes a fixed energy bias and $\kappa$ denotes the measurement rate.

To train the neural network, we prepare a dataset consisting of several initial states of the two-level system that are then evolved using the \textbf{smesolve} method from the QuTIP python package used for simulating open quantum systems\cite{JOHANSSON20131234}. %
We then train our transformer using the data set consisting of a range of initial states, their associated measurement records, and a locally optimal control protocol, $\lambda_t$, that drives the system to the target state for each noise realization. This locally optimal control is found using the PaQS algorithm~\cite{Zhang2020}. During the training phase, we always set $\varepsilon=0$.

\begin{figure}
    \centering
    \includegraphics[width=0.5\textwidth]{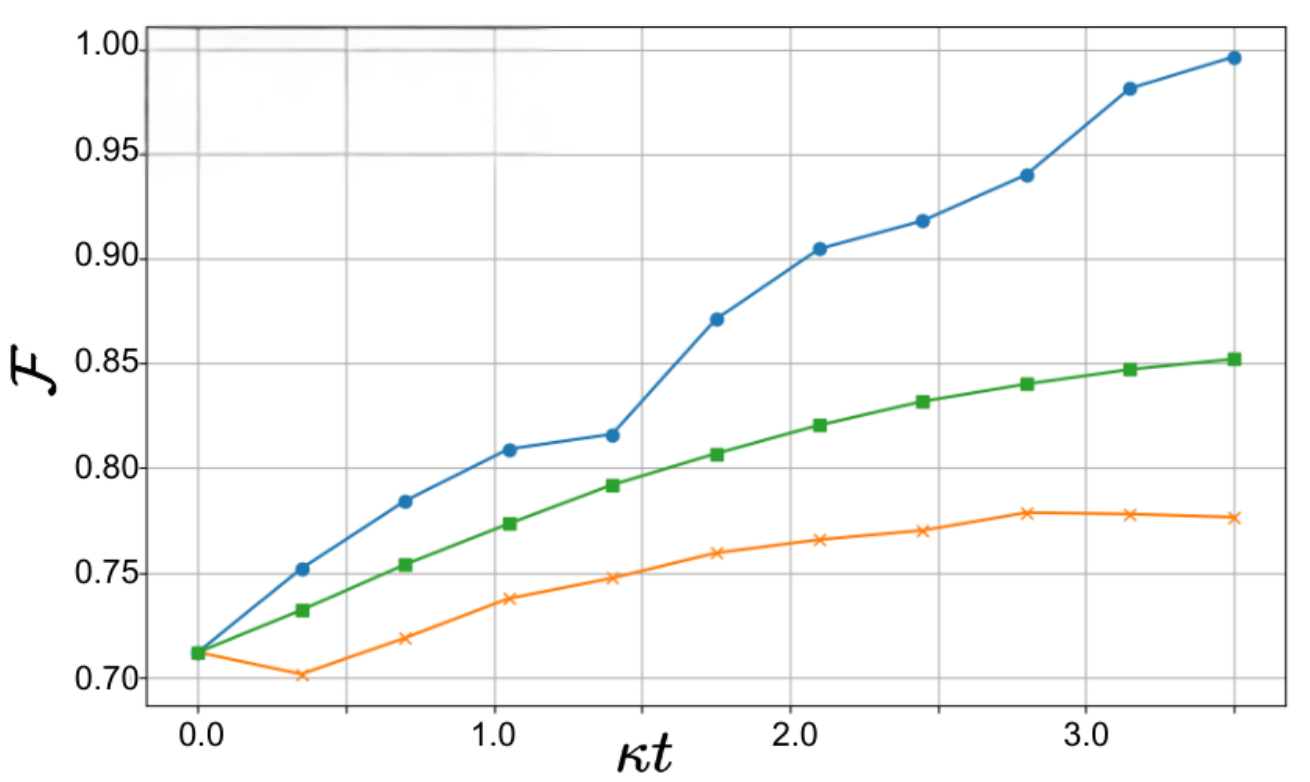}
    \caption{Fidelity $\mathcal{F}$ with a target state as a function of time under feedback control. The initial state is $\hat{\rho}_0 = |\psi_0\rangle \langle \psi_0|$, where $\ket{\psi_0} =\alpha|0\rangle+\beta|1\rangle$ with $\alpha = \sqrt{\frac{7}{12}}$ and $\beta= \sqrt{\frac{5}{12}}$. The target state is $\ket{\psi_{\rm targ}}=\frac{1}{\sqrt{2}}(|0\rangle+i|1\rangle)$. The performance of the transformer under imperfect measurement efficiency ($\eta=0.7$) (blue, circles), and an increase in bias ($\epsilon =0.5$) (green, squares) is benchmarked against the fidelity improvement when randomly selecting $\lambda_t$ values (orange, crosses).}
    \label{fig:fidelity_TLS}
\end{figure}

As seen in Fig. \ref{fig:fidelity_TLS}, the transformer generates feedback strategies that can stabilize the TLS in a coherent superposition state $\ket{\psi_{\rm targ}} = (\ket{0}+i\ket{1})/\sqrt{2}$ even with inefficient measurements (we take $\eta =0.7$). The transformer approach is also robust against perturbations of the dynamics, as we show by introducing a significant bias $\varepsilon\neq 0$, which was absent during training. Further examples can be found in Supplemental Material \cite{supplement}.

Another beneficial feature of the transformer is the speed with which it outputs the optimal control parameter for the next time step. Table~\ref{tab:my_label} compares the time taken to infer the entire trajectory by the transformer algorithm and the modified proportional and quantum state (PaQS) algorithm, where the latter requires solving the stochastic master equation~\eqref{SME} at each time step. We observe a speed-up of approximately two orders of magnitude in our numerical tests, which were performed using a standard laptop. This is because PaQS, and other iterative numerical solvers require multiple evaluations of the right-hand side of Eq. \eqref{SME} per gradient evaluation and per line-search, whereas our deployed transformer avoids online integration of the stochastic master equation altogether. However, it should be noted that this speed advantage comes at the cost of a large memory required to store the neural network representation. This memory requirement is likely to prove the most significant bottleneck for integrating the transformer into optimized hardware such as GPUs or FPGAs \cite{Reuer2023}.

\begin{table}[b]
\begin{ruledtabular}
\begin{tabular}{ll}
Method                        & Inference Time (in sec) \\
Hamiltonian Modified PaQS     & 19.05                       \\
Quantum Transformer Inference & 0.23             
\end{tabular}
\end{ruledtabular}
    \caption{The inference time in seconds of predicting optimal $\lambda_t$ for a single trajectory with 100 discretized time steps during the evolution of the state governed by the stochastic master equation(\ref{SME}). We benchmark the inference time of the transformer against the time taken to calculate the optimal feedback operation using a gradient based solver for the PaQS approach. The inference is run on a 2021 Macbook Pro with 16GB of RAM and a 8-core CPU.}
    \label{tab:my_label}
\end{table}

In order to provide another example to demonstrate the flexibility and generalizability of our transformer-based approach, we apply it to the challenging problem of controlling non-Markovian quantum dynamics~\cite{li2020non}. Specifically, we now consider our TLS to be coupled to a harmonic oscillator mode with angular frequency $\Omega$ and coupling strength $g$, leading to the Hamiltonian
\begin{equation}
	\label{RC_Hamiltonian}
	\hat{H}(\lambda_t) =\frac{\hbar \varepsilon}{2} \hat{\sigma}_z+\frac{\hbar \lambda_t}{2} \hat{\sigma}_x + \hbar\Omega \hat{a}^\dagger \hat{a} +\hbar g \hat{\sigma}_z(\hat{a}+\hat{a}^\dagger). 
\end{equation}
We assume that the oscillator mode is coupled to a broadband environment that is continuously monitored via homodyne detection, leading to a stochastic master equation of the form~\eqref{SME} with $\hat{c}=\sqrt{\kappa} \hat{a}$. This situation can be realized, for example, by a superconducting qubit interfaced with a cavity resonator that is itself coupled to a waveguide~\cite{Blais2004}. This situation is well known to lead to non-Markovian dynamics for the qubit when the cavity linewidth $\kappa$ is not too large, i.e. if $\kappa\lesssim g$~\cite{breuer2002theory}.

Alternatively, one can interpret the cavity mode as a ``reaction coordinate'' (RC), which represents a collective mode of a structured reservoir whose spectral density is peaked at frequency $\Omega$~\cite{IlesSmith2014,tamascelli2018nonperturbative}. The extended open quantum system comprising the TLS and RC can thus be understood as a Markovian embedding of the original non-Markovian dynamics induced by the structured reservoir~\cite{Woods2014}. Meanwhile, the residual (broadband) environment represents far-field degrees of freedom that can be monitored without disrupting the non-Markovian character of the TLS evolution.

\begin{figure} 

\centering

	\includegraphics[width=0.50\textwidth]{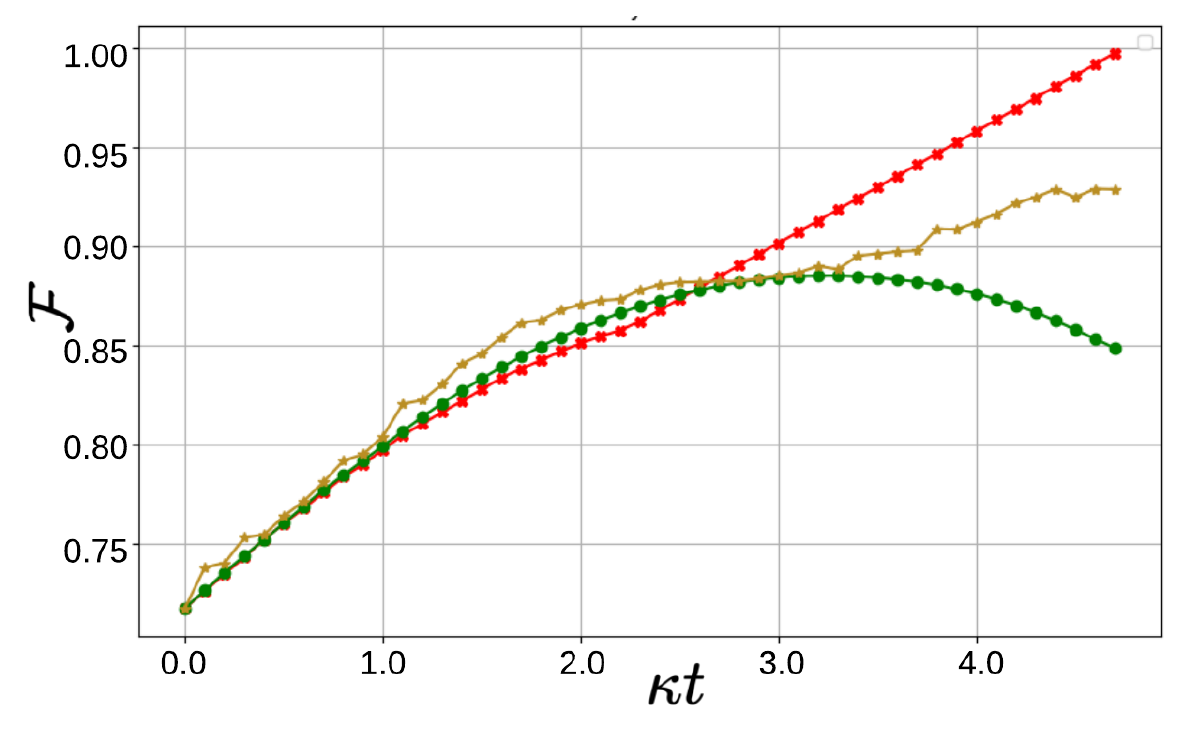}
	\caption{The fidelity $\mathcal{F}$ as a function of time while benchmarking the performance of the transformer (red, crosses) as compared to a vanilla recurrent neural network (green, circles) and a gated-recurrent unit recurrent neural network GRU-RNN (yellow, stars). The context of 2000 measurement record samples is provided in the case of the non-Markovian setting accounted for by the reaction coordinate embedding. The coupling with the bath provided by $g=0.5$. The dimension of the reaction coordinate is truncated to $d=6$.}
	\label{rnn-transformer}
\end{figure}

Since non-Markovian effects can affect the system dynamics in a much longer time horizon, the attention-based transformer model seems ideal to control non-Markovian systems due to the self and multihead attention. We use transfer learning to fine-tune the transformer to predict optimal $\lambda_t$ values for the reaction-coordinate setting with a smaller dataset. As seen in Fig. \ref{rnn-transformer}, the transformer learns to predict optimal $\lambda_t$ even in this non-Markovian setting. To benchmark the transformer, we train a vanilla recurrent neural network (RNN) and a gated recurrent unit recurrent neural network (GRU-RNN) with up to 60 time steps of the measurement record \cite{cho2014learningphraserepresentationsusing}. We train the RNNs on the given number of time steps to avoid the vanishing gradient problem as explored in the literature of deep learning \cite{pmlr-v28-pascanu13}. We can observe that even though the vanilla RNN and GRU-RNN perform slightly better than the transformer for shorter time periods, where the measurement record provided is much smaller, the transformer outperforms the vanilla RNN and the GRU-RNN in the case of long context windows and non-Markovian closed-loop feedback control. This is likely due to the fact that the transformer can process and attend to arbitrary measurement record lengths and does not suffer from the requirement of a sequential approach such as from RNNs.

\textit{Many-body state preparation.---}In the preceding examples, we demonstrated that our transformer can be trained via supervised learning, 
using known locally optimal protocols such as PaQS. We now turn to the example of approximate ground-state preparation of a many-body system via feedback control \cite{metz2023self}, where no such baseline is available and our transformer can no longer be trained in a supervised manner \cite{Zhang_2023}. Consider an open $N$‑qubit mixed‑field Ising chain governed by the Hamiltonian
\begin{equation}
\label{Ising_Hamiltonian}
    \hat{H}(\lambda_t)=\lambda_t \sum_{i=1}^N \hat{\sigma}_i^z+g \sum_{i=1}^{N-1} \hat{\sigma}_i^x \hat{\sigma}_{i+1}^x+h\sum_{i=1}^N \hat{\sigma}_i^x,
\end{equation}
where $g$ is the nearest-neighbor coupling strength, $h$ is a static longitudinal field, and the transverse field $\lambda_t$ is the control parameter. Our goal is to sweep this parameter from an initial value $\lambda_0 \approx 0$---for which the tensor-product ground state is generally easy to prepare---to a final value $\lambda_T \gtrsim g$, within a fixed time $T$, while ideally leaving the system close to the ground state of the final Hamiltonian. This is generally challenging in a finite-time protocol due to the creation of non-adiabatic excitations~\cite{Kolodrubetz2017,Odelin2019}. Moreover, the Hamiltonian in Eq.~\eqref{Ising_Hamiltonian} is non-integrable~\cite{Kim2013,Nieva2024} and thus finding locally optimal control solutions becomes unfeasible as $N$ grows larger.

The feedback is now conditioned on a homodyne measurement of the collective spin operator $\hat{c} = \sqrt{\kappa}\sum_i \hat{\sigma}_i^z$, which produces diffusive dynamics as given by Eq.~\eqref{SME}, and an additional measurement of the mean energy ${\rm tr}[ \hat{H}(\lambda_T)\hat{\rho}_T]$ at the final time. The latter provides the loss function to be minimised by the control protocol for a fixed time $T$ and endpoints $\{\lambda_0,\lambda_T$\}. Crucially, since labeled optimal trajectories are now absent, we utilize a model-free reinforcement learning type approach \cite{janner2021offline}, that allows our model to learn optimal control pulses through trial and error by direct interaction and observation of rewards. We use the strategy of an online iteratively-refined decision transformer (IR-DT) training mechanism \cite{zheng2022online}. Our inputs remain consistent with the previous examples; however, we also include the return-to-go award based on the final energy. Table \ref{tab:spin-energies} shows the mean final energies reached by our trained transformer, demonstrating good performance even as the system size grows, e.g.~we observe a maximum discrepancy with the true ground-state energy of $~2\%$ for 8 spins. These results demonstrate the flexibility of our transformer architecture, as it can be adapted to situations where model-based supervised learning is challenging or impossible.

\begin{table}
  \centering
  \begin{tabular}{|c|c|c|}
    \hline
    Spins & Ground State Energy & Mean Predicted Energy \\
    \hline
    4 & -6.518$g$ & -6.209$g$ \\
    6 & -9.723$g$ & -9.666$g$ \\
    8 & -13.155$g$ & -12.845$g$ \\
    \hline
  \end{tabular}
  \caption{Comparison of ground-state and predicted energies for different number of spins. We set $g=6.07$, $\lambda_0=-0.097g$, $\lambda_T=1.27g$, $h=1.19g$, $\kappa=0.148g$ and $gT=5.56$.}
  \label{tab:spin-energies}
\end{table}

\textit{Conclusion-} In this work, we have presented a novel approach for closed-loop adaptive feedback control of open quantum systems using attention-based transformer neural networks. We have demonstrated that our quantum transformer model can effectively learn to predict optimal control parameters based on the initial state and measurement record of a two-level quantum system. The transformer architecture, with its self-attention mechanism, enables capturing long-range dependencies in the measurement record, outperforming traditional methods like recurrent neural networks for feedback control in non-Markovian systems where different temporal aspects of the measurement record may affect the state evolution. Due to these reasons, transformer models are an ideal candidate for applications in quantum prediction \cite{rodriguez2024shorttrajectoryneedtransformerbased} as well as control which we have demonstrated in this work.\par

We have shown the robustness and scalability of our approach under various conditions, such as imperfect measurement efficiency and perturbations in the Hamiltonian. Furthermore, using transfer learning, we have successfully applied our transformer model to the challenging task of controlling non-Markovian open quantum systems, which is a significant advancement in the field. The attention-based approach to quantum feedback control offers several advantages, including faster inference times compared to state-of-the-art methods like the Hamiltonian Modified PaQS. The transformer's ability to handle long context windows and its scalability make it a promising tool for controlling complex quantum systems. Furthermore, we demonstrated the transformer's ability to perform reinforcement learning when optimal parameters for supervised training are inaccessible, highlighting the promise of our approach for tackling challenging many-body control problems. In conclusion, our work demonstrates the effectiveness of attention-based transformer models for closed-loop feedback control of Markovian, non-Markovian, and many-body open quantum systems. This approach opens up new possibilities for the development of robust and efficient quantum control techniques, which are crucial for the advancement of quantum technologies. Future work could explore the application of this method to even more complex quantum systems and investigate its performance in experimental settings.

\textit{Acknowledgements.} The authors acknowledge useful discussions with Prof. Gerard Milburn. N.A. acknowledges support from the European Research Council (grant agreement 948932) and the Royal Society (URF-R1-191150). M.T.M. is supported by a Royal Society University Research Fellowship. The research of F.M. is partially supported by the Munich Quantum Valley, which is supported by the Bavarian state government with funds from the Hightech Agenda Bayern Plus.  This project is co-funded by the European Union and UK Research \& Innovation (Quantum Flagship project ASPECTS, Grant Agreement No.~101080167). Views and opinions expressed are however those of the authors only and do not necessarily reflect those of the European Union, Research Executive Agency or UK Research \& Innovation. Neither the European Union nor UK Research \& Innovation can be held responsible for them. P.V. is supported by the United States Army Research Office under Award No. W911NF-21-S-0009-2. The authors would like to acknowledge the use of the University of Oxford Advanced Research Computing (ARC) facility in carrying out this work. http://dx.doi.org/10.5281/zenodo.22558.

\textit{Data Availability.} The data associated with model architecture, training and data generation are publicly available \cite{github_repo}. Due to the size of the dataset generated for training and the hosting constraints, the generated data is available upon request.

\bibliographystyle{apsrev4-2}
\bibliography{bibliography}%
\appendix

\section{Appendix A: Local Optimality Using the PaQS Approach}

In this appendix, we summarize the PaQS approach of Zhang \textit{et al.}~\cite{Zhang2020} as applied to our problem of interest. For simplicity, we set the bias $\varepsilon=0$ and focus on the effect of varying $\lambda_t$ to control the system. We  consider an arbitary target state: 
\begin{equation}\label{target_state}
    \left|\psi_T\right\rangle=\left|\psi_{\text {system }}\right\rangle \otimes\left|\psi_{\mathrm{RC}}\right\rangle 
\end{equation}
for some generic state $\left|\psi_{\text{system}}\right\rangle$ and $\left|\psi_{\mathrm{RC}}\right\rangle$ of the reaction coordinate. As explained in the main text, the role of the reaction coordinate is essential for performing the Markovian embedding of a non-Markovian system. When solving the optimal control of the system, the combined state of the system and reaction coordinate is treated collectively in the density matrix, allowing standard Markovian master equation techniques to be applied to this extended system. However, it makes the analysis somewhat more complicated than Ref.~\cite{Zhang2020} because of the presence of the Hamiltonian of the reaction coordinate, in addition to the control Hamiltonian proportional to $\lambda_t$. 

To proceed, we separate the evolution with and without feedback. In the absence of feedback, the state evolves over a timestep $dt$ according to the stochastic master equation ($\hbar=1$)
\begin{equation}
    \label{SME_0}
    d \hat{\rho}_t =-i [\hat{H}_0, \hat{\rho}_t]dt + \mathcal{D}[\hat{c}] \hat{\rho}_t dt + \sqrt{\eta} \mathcal{H}[\hat{c}] \hat{\rho}_t d W_t,
\end{equation}
with the Hamiltonian $\hat{H}_0 \equiv \hat{H}(\lambda_t = 0)$. Following Zhang \textit{et al.}, we describe the feedback control by the unitary operator
\begin{equation}\label{feedback}
    \hat{U}(\theta_t) \equiv e^{-i \theta_t \hat{H}_F},
\end{equation}
where $\hat{H}_F = \sigma_x/2$ is the feedback Hamiltonian and the infinitesimal rotation angle $\theta_t = \lambda_t dt$ encapsulates the effect of the control parameter. In the following analysis, for the sake of simplicity, we consider the measurement efficiency $\eta =1$. The fidelity with respect to a target state $|\psi_T\rangle$ is $\mathcal{F}_t = \left\langle\psi_T\left|\hat{\rho}_t\right| \psi_T\right\rangle$, which thus updates according to
\begin{equation}
\label{fidelity_update}
\mathcal{F}_{t+d t}  =\langle\psi_T|\hat{U}(\theta_t)\left[\hat{\rho}_t+d \hat{\rho}_t\right] \hat{U}^{\dagger}(\theta_t)| \psi_T\rangle. \end{equation}

The goal of locally optimal control is to choose the rotation angle $\theta_t$ to maximize the fidelity with the target state at each step. We therefore demand $\mathcal{G} = 0$, where the cost function is
\begin{equation}\label{cost}
    \mathcal{G} \equiv \frac{\partial \mathcal{F}_{t+d t}}{\partial \theta_t}=-i\left\langle\psi_T\left|\left[
\hat{H}_F, \hat{\rho}_t\right]\right| \psi_T\right\rangle + \mathcal{O}(dt),
\end{equation}
and we keep only the leading-order term, neglecting infinitesimal corrections. We denote by $\theta^*_t$ the optimal value of $\theta_t$ that solves $\mathcal{G} = 0$. Since $\theta_t^*$ is infinitesimal, it can be parameterized without loss of generality as~\cite{Zhang2020}
\begin{equation}
    \theta_t^*=A_1(t) d W_t +A_2(t) d t,
\end{equation}
where $A_1$ and $A_2$ are to be solved for. 

To get explicit expressions for the functions $A_1$ and $A_2$, we expand the unitary operator $U_F$ up to second order in $dW_t$ and make use of the rules of Ito calculus, i.e., $dW^2 = dt$ and $dW dt = 0 = dt^2$. The second-order expansion of $U(\theta^*_t)$ is given by
\begin{equation}
U= I-i A_1 \hat{H}_F d W-\left(i A_2 \hat{H}_F+\frac{1}{2} A_1^2 \hat{H}_F^2\right) d t.
\end{equation}
Substituting this into the state update rule $\hat{\rho}_{t+dt} = \hat{U}(\theta^*_t)[\hat{\rho}_t+d\hat{\rho}_t]\hat{U}^\dagger(\theta^*_t)$ and simplifying yields
\begin{widetext}
    \begin{equation}
    \begin{aligned}
\hat{\rho}_{t+dt}= & \hat{\rho}_t+d t\left(-i\left[\hat{H}_0, \hat{\rho}_t\right]+\kappa \mathcal{D}[\hat{a}] \rho-i A_2\left[\hat{H}_F, \hat{\rho}_t\right]-\frac{1}{2} A_1^2\left \{\hat{H}_F^2, \hat{\rho}_t\right\}\right. \\
& \left.+A_1^2 \hat{H}_F \hat{\rho}_t \hat{H}_F-i A_1 \sqrt\kappa\left[\hat{H}_F, \mathcal{H}[\hat{a}] \hat{\rho}_t\right]\right) \\
& d W_t\left(-i A_1\left[\hat{H}_F, \hat{\rho}_t\right]+\sqrt{\kappa} \mathcal{H}[\hat{a}] \hat{\rho}_t\right). \\
&
\end{aligned}
\end{equation}
We can now substitute this expression into the cost function in Eq. \eqref{cost} to find
\begin{equation} \label{eq:costfunction}
  \mathcal{G} =  -i\left\langle\psi_T\left|\begin{array}{r}
\hat{\rho}_t-i A_2 d t\left[\hat{H}_F, \hat{\rho}_t\right]-i A_1 d W\left[\hat{H}_F, \hat{\rho}_t\right]+d t\left(-i\left[\hat{H}_0, \hat{\rho}_t\right]+\kappa \mathcal{D}[a] \hat{\rho}_t+A_1^2 \mathcal{D}\left[\hat{H}_F\right] \hat{\rho}_t\right. \\
\left.-i A_1 \sqrt{\kappa}\left[\hat{H}_F, \mathcal{H}[a] \hat{\rho}_t\right]\right)+\sqrt{\kappa} \mathcal{H}[a] \hat{\rho}_t d W
\end{array}\right| \psi_T\right\rangle = 0.
\end{equation}
Since this cost is defined for the target state as mentioned in Eq. (\ref{target_state}) which consists of $\left|\psi_{\mathrm{RC}}\right\rangle$. We can then solve Eq. (\ref{eq:costfunction}) in a truncated Hilbert space numerically using algorithms such as the Broyden-Fletcher-Goldfarb-Shanno (BFGS) or the modified Newton-Raphson algorithm to find the optimal $A_1$ and $A_2$ \cite{wright2006numerical}. Following a similar analysis to solving $\mathcal{G}$ analytically, we get:
\begin{equation}
A_1 = \frac{\sqrt{\kappa}\left\langle\psi_T\left|\mathcal{H}[\hat{a}] \hat{\rho}_t\right| \psi_T\right\rangle}{i\left\langle\psi_T\left|\left[\hat{H}_F, \hat{\rho}_t\right]\right| \psi_T\right\rangle}
\end{equation}
and 
\begin{equation}
    A_2 = \frac{\left\langle\psi_T\left|\left(-i\left[\hat{H}_0, \hat{\rho}_t\right]+\kappa \mathcal{D}[\hat{a}] \hat{\rho}_t+A_1^2 \mathcal{D}\left[\hat{H}_F\right] \hat{\rho}_t - i A_1 \sqrt{\kappa}\left[\hat{H}_F, \mathcal{H}[\hat{a}] \hat{\rho}_t\right]\right)\right| \psi_T\right\rangle}{i\left\langle\psi_T\left|\left[\hat{H}_F, \hat{\rho}_t\right]\right| \psi_T\right\rangle}
\end{equation}
\end{widetext}

\begin{figure*}
\includegraphics[width=\textwidth]{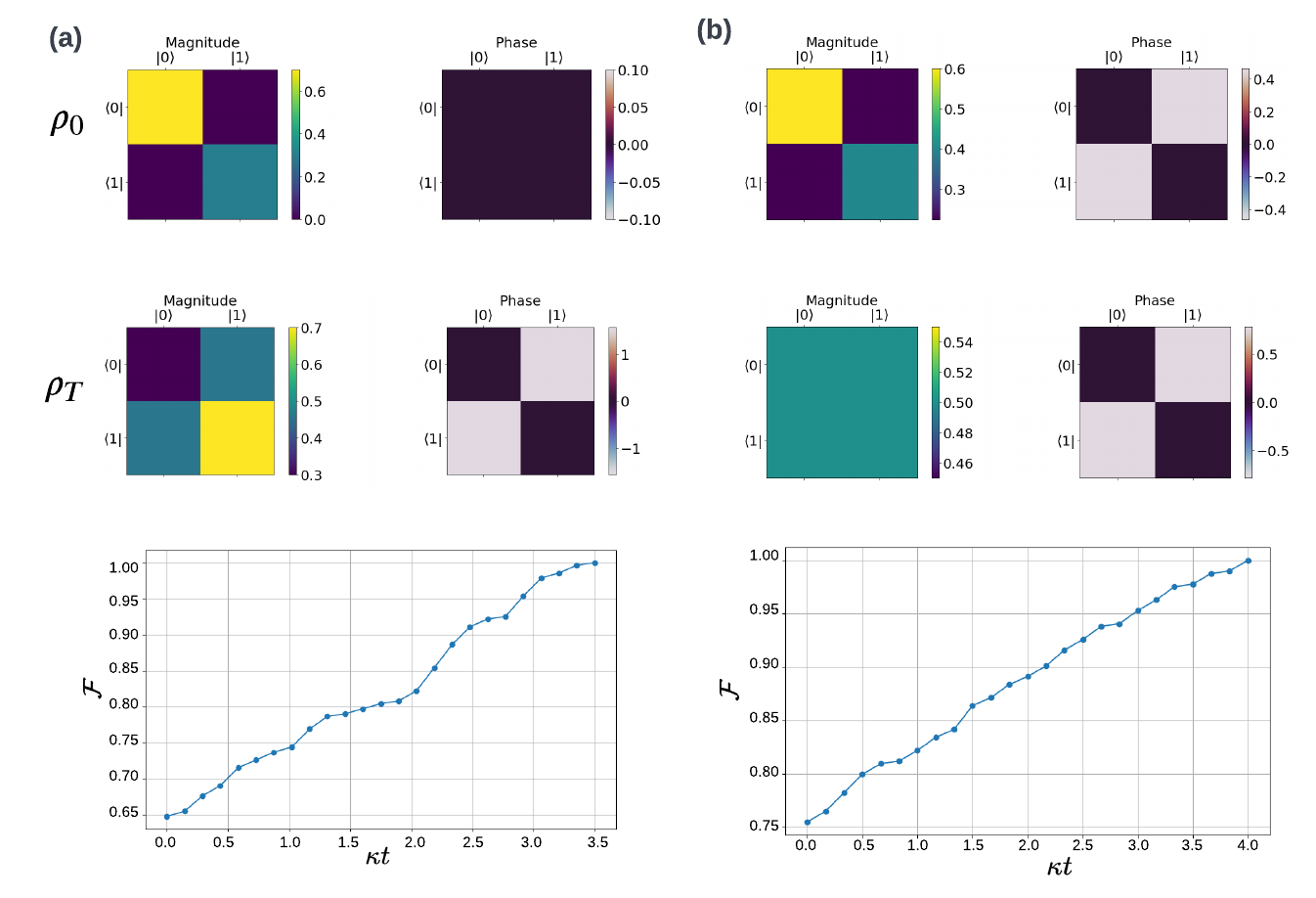}
    \caption{\label{fig:appendix}\textbf{a,} The magnitude and phase map of the mixed initial state of $\rho_0=0.7|0\rangle\langle 0|+0.3|1\rangle\langle 1|$ and target pure state of $\left|\psi_T\right\rangle=\sqrt{0.3}|0\rangle+i \sqrt{0.7}|1\rangle$. The third plot represents the increase in fidelity towards the target pure state based on control parameters produced by the transformer (blue, circles) when undergoing continuous measurement with a measurement efficiency of 0.8. \textbf{b,} The magnitude and phase map of the mixed initial state of $\rho_0=0.6|0\rangle\langle 0|+(0.2+0.1 i)|0\rangle\langle 1|+(0.2-0.1 i)|1\rangle\langle 0|+0.4|1\rangle\langle 1|$ and target pure state of $\psi_T=\frac{|0\rangle}{\sqrt{2}}+\frac{1+i}{2}|1\rangle$. The third plot represents the increase in fidelity towards the target pure state based on control parameters produced by the transformer (blue, circles) when undergoing continuous measurement with a measurement efficiency of 0.65.} 
    \label{appendix}
\end{figure*}

\section{\label{app:Section} Appendix B: Attention Formalism}

Transformers are inherently permutation-invariant due to their parallel processing of input sequences, which poses a challenge when dealing with sequential data where order is crucial. To address this, we incorporate positional embeddings into the measurement record input. Positional embeddings encode temporal relationships between different parts of the measurement record, allowing the model to capture the sequence's temporal dynamics. This encoding ensures that the model can distinguish between measurements taken at different time steps, which is essential for accurately predicting the optimal control parameters over time \cite{vaswani2017attention}.

The encoder's output serves as a context vector summarizing the initial state and measurement record information. The self-attention mechanism can be conceptualized as a graph-like structure, where each element in the input sequence is connected to every other element. The strength of these connections, or attention weights, is learned through training. This allows the model to weigh the importance of different parts of the input when generating an output. Mathematically, the self-attention mechanism can be described as a weighted sum of value vectors, where the weights are determined by the compatibility between query and key vectors. These vectors are obtained by applying learned linear transformations to the input embeddings. The QuantumDecoder module is structured similarly to a generative pretrained transformer (GPT) module along with positional embeddings to perform an autoregressive task \cite{radford2018improving}. During training, the QuantumDecoder module --- which consists of embedding layers for the optimal parameter values and measurement record, followed by a Transformer decoder --- takes the optimal parameter values, context (encoded representation from the encoder), and measurement record as input and predicts the next value $\lambda_t$ in the sequence.

In our transformer-based model for quantum control, the attention mechanism is important to capture the complex dependencies between the initial quantum state, the sequential measurement records, and the optimal control parameters that we aim to predict. Here, we focus on a detailed, yet concise explanation of how attention operates within our model, considering both self-attention and cross-attention. The cross-attention layers in the decoder enable it to integrate contextual information from the encoder.

Cross-attention bridges the encoder and decoder by computing attention scores between the decoder's queries and the encoder's keys and values. Several attention scores can be calculated by different linear transformation known as attention heads. Multiple attention heads operate in parallel, and their outputs are concatenated and transformed. Feed-forward networks introduce non-linearity and residual connections and layer normalization are applied for stable training. We use multiple encoder modules together to learn hierarchical representations as well. By stacking several identical Transformer-encoder blocks—each comprising multi-head self-attention, a feed-forward network, and residual normalization—the model builds depth that progressively abstracts the input. Early layers focus on very local measurement correlations, mid-level layers capture medium-range dependencies, and the deepest layers integrate global structure needed for long-horizon planning. This leads to better out-of-distribution generalization as well.

While reinforcement learning methods can learn control strategies through interaction with the environment, they often require extensive exploration and can suffer from instability during training. Our transformer-based approach provides a stable and efficient alternative by learning from supervised data. It eliminates the need for exploration by utilizing known optimal control parameters during training. This approach leverages the strengths of sequence modeling inherent in transformers, capturing long-range dependencies without the high variance typically associated with RL methods \cite{li2023surveytransformersreinforcementlearning}.

Self-attention allows the model to weigh the relevance of different elements within a single input sequence by computing attention scores between all pairs of positions. For each element in the sequence, the model generates query vectors ($Q$), key vector ($K$), and value vector ($V$) using learned linear transformations.
\begin{equation}
\begin{aligned}
Q & =W_Q \cdot X \\
K & =W_K \cdot X \\
V & =W_V \cdot X
\end{aligned}
\end{equation}
where $X$ is the latent space embeddings from the input and $W_Q$, $W_K$ and $W_V$ are learned weight vectors. The self attention can be calculated as:
\begin{equation}
	\operatorname{Attention}(Q, K, V)=\operatorname{softmax}\left(\frac{Q K^T}{\sqrt{d_k}}\right) V.
\end{equation}
When we include both the initial quantum state and the measurement record in the encoder, self-attention computes how each element (state or measurement at a certain time) relates to every other element in the sequence. This enables the encoder to build a comprehensive contextual representation that captures relationships between the initial conditions and the observed dynamics.

The decoder processes the measurement record up to the current time step. Masked self-attention ensures that predictions for the control parameters at each time step only depend on current and past measurements, preserving causality. This mechanism allows the decoder to understand temporal dependencies within the measurement sequence.

The cross-attention occurs in the decoder and allows it to incorporate information from the encoder's output. The decoder's queries attend to the encoder's keys and values, integrating contextual information from the encoder:
\begin{equation}
\operatorname{Attention}=\operatorname{softmax}\left(\frac{Q^{\mathrm{dec}}\left(K^{\mathrm{enc}}\right)^{\top}}{\sqrt{d_k}}\right) V^{\mathrm{enc}}.
\end{equation}

\section{Appendix C: Model and Dataset Details}
The model consists of 6 encoder and 6 decoder layers. Each layer comprises an embedding dimension of 512 and 8 attention heads. In order to maximize performance, a context window of 1024 tokens was chosen. Training was performed using the RAdam optimizer with an initial learning rate of $0.001$ and was trained for 100 epochs with early stopping on validation loss to prevent overfitting \cite{liu2021varianceadaptivelearningrate}.  Gradient clipping with a maximum norm of 1 was applied to stabilize training and prevent exploding gradients. Model training was performed on a NVIDIA A100 GPU with 80GB of RAM.

The dataset was generated using the QuTiP package~\cite{JOHANSSON20131234}. There were 200 unique initial quantum states generated. For each initial state, the smesolve method was used to simulate 1,000 stochastic trajectories using the stochastic master equation, with each trajectory consisting of 1,000 time steps.

\section{Appendix D: Examples}

In this appendix, we present more examples to demonstrate the performance of the transformer further. We provide examples of state preparation and purification from a varied set of initial and target states as seen in Fig \ref{appendix}. We also demonstrate that the transformer predicts optimal control parameters for state purification and preparation even under measurement inefficiencies under continuous measurement. 

\section{Appendix E: Iteratively Refined Decision Transformer(IR-DT)}

The architecture we use for the reinforcement learning task of learning the ground state energy starts from a conventional Decision-Transformer. Because no expert demonstrations exist for the transverse-Ising task, we first \textit{pre-train} the network on a modest offline buffer of random control strings together with the energies they realize .  Then an outer self-improvement loop is launched following these steps:

\begin{itemize}
    \item Roll out the current transformer policy to generate $M$ new trajectories.
    \item Re-label each trajectory with the actual energy return obtained at the end of the sweep.
    \item Aggregate these fresh state-action-return triples into the replay buffer.
    \item Fine-tune the same network on the enlarged dataset.
\end{itemize}

Because the Decision-Transformer conditions every decoder step on the desired return, newly added low-energy examples automatically bias the next policy towards deeper minima while maintaining training stability.   Causal masking enforces autoregressive conditioning, while return conditioning biases the output toward lower energies without recourse to a separate value network or policy gradient as seen in traditional reinforcement learning methods \cite{schulman2017proximalpolicyoptimizationalgorithms}. Repeating this refine-and-retrain cycle drives the trajectory distribution monotonically downward in energy without ever computing policy gradients or value functions, allowing the transformer to discover near-ground-state control schedules for chains up to $N=8$. Note, however, that this approach is extremely data intensive requiring more than $>100k$ trajectories.
\end{document}